\documentclass{elsart}
\usepackage{amssymb,amsfonts} %
\usepackage[dvips,
            ]{hyperref} 
\def\cal{\mathcal}

\journal{arXiv.org}
\begin{document}
\begin{frontmatter}
\title{$Z_4$-linear Hadamard and extended perfect codes\thanksref{zzzz}}
\author{
{D.\,S.\,Krotov}}
\thanks[zzzz]{This is a reprint of: --''--
 In {\em Proc. the {I}nt. Workshop on Coding and Cryptography
  \href{http://www-rocq.inria.fr/codes/WCC2001/}{{WCC}'2001}}, pages 329--334,
  Paris, France, January 2001.
 \emph{Electron. Notes Discrete Math.}, 6: 107--112, 2001.
  \href{http://dx.doi.org/10.1016/S1571-0653(04)00161-1}{DOI: 10.1016/S1571-0653(04)00161-1}}

\ead{krotov@math.nsc.ru}
\address{Sobolev Institute of Mathematics, Russia}

\begin{abstract}
If $N=2^k\geq 16$ then
there exist exactly
$\lfloor (k-1)/2\rfloor$ pairwise nonequivalent $Z_4$-linear
Hadamard $(N,2N,N/2)$-codes and
$\lfloor (k+1)/2\rfloor$ pairwise nonequivalent $Z_4$-linear
extended perfect $(N,2^N/2N,4)$-codes. A recurrent construction
of $Z_4$-linear Hadamard codes is given.
\end{abstract}
\begin{keyword}
Hadamard Codes \sep Perfect Codes \sep $Z_4$-Linear Codes
\end{keyword}
\end{frontmatter}

\section{Introduction}

Certain of known nonlinear binary codes such as Kerdock, Preparata, Goethals,
Delsarte-Goethals codes are represented
by use of a map $\{0,1,2,3\}\to \{0,1\}^2$
as linear codes over the alphabet $\{0,1,2,3\}$ with modulo $4$ operations
(see \cite{Nech,Z4})
(following \cite{Z4}, we will use the map $0\to 00$, $1\to 01$, $2\to 11$, $3\to 10$).
Codes represented in such a manner are called $Z_4$-linear.

Our  research is devoted to
$Z_4$-linear Hadamard $(N,2N,N/2)$-codes
and $Z_4$-linear extended perfect
$(N,2^N/2N,4)$-codes.
Linear in the ordinary sense \linebreak $(N,2N,N/2)$-code and $(N,2^N/2N,4)$-code
exist for every $N=2^k$ and unique up to equivalence.
These codes are first order Reed-Muller code and extended Hamming code
respectively.
In \cite{Z4} it was shown that the first order Reed-Muller codes
are $Z_4$-linear and the Hamming code of length $N$ is
$Z_4$-linear if and only if $N\leq 16$.
Also in \cite{Z4} a $Z_4$-linear
$(N,2^N/2N,4)$-code
was presented in a cyclic form for every $N=2^k$.
The aim of our research is a full up to equivalence classification of
$Z_4$-linear $(N,2N,N/2)$- and $(N,2^N/2N,4)$-codes.
The results on extended perfect $(N,2^N/2N,4)$-codes are proved in \cite{Kr}.
(A complete classification of the $Z_4$-linear Hadamard codes can be found in \cite{PheRifVil:2006};
for more references on the subject, see \href{http://arxiv.org/abs/0710.0198}{arXiv:0710.0198}
-- transl. rem.)

\section{Main definitions and facts}

Let $E^N$ be the set of all binary words of length $N$.
{\it Hamming distance} $d(x,y)$ between $x$ and $y$ from $E^N$ is
the number of positions in which $x$ and $y$ differ.
Binary {\it $(N,K,d)$-code } is a subset $C$ of $E^N$
such that $|C|=K$ and $d(c_1,c_2)\geq d$ for every different $c_1,c_2\in C$.
If $c_1\oplus c_2\in C$ for every $c_1,c_2\in C$ then
 $C$ is {\it linear} code.

Let $Z_4^n$ be the set of $n$-words over the alphabet $Z_4=\{0,1,2,3\}$
with (mod\,$4$) addition and multiplication by a constant.
An additive subgroup of $Z_4^n$ is called a {\it quaternary code}.
Two quaternary codes are {\it equivalent} if
one can be obtained from the other by permuting the coordinates and
(if necessary) changing the signs of certain coordinates.

{\it Lee weight} $wt_L(a)$ of $a\in Z_4^n$ is the rational sum
of the Lee weights of its coordinates, where
$wt_L(0)=0$, $wt_L(1)=wt_L(3)=1$ and $wt_L(2)=2$.
The weight function $wt_L$ defines {\it Lee distance}
$d_L(a,b)\stackrel{\rm\scriptscriptstyle def}=wt_L(b-a)$ on $Z_4^n$.

We say that a quaternary code
${\cal C}$ is a {\it quaternary distance $d$ code of length $n$}
or ${\cal C}$ is a {\it $(n,|{\cal C}|,d)_4$-code} if ${\cal C}\subseteq Z_4^n$
and  $d_L(a,b)\geq d$ for every different $a,b\in\cal C$.

Every quaternary code $\cal C$ can be defined by a {\it generating matrix} of the form
\begin{equation}{G=\left[\matrix{G_1\cr 2G_2}\right],\label{gen}}
\end{equation}
where $G_1$ is a $Z_4$-matrix of size $k_1\times n$,
$G_2$ is a $Z_2$-matrix of size $k_2\times n$,
$|{\cal C}|=2^{2k_1+k_2}$, and every $c\in\cal C$ can be represented in form
\[c=(v_1,v_2)\left[\matrix{G_1\cr 2G_2}\right]\mbox{ (mod 4)},
\qquad v_1\in Z_4^{k_1},\quad v_2\in Z_2^{k_2}.\]

The code $\cal C$ defined by generating matrix (\ref{gen}) is
an elementary Abelian group of type $4^{k_1}2^{k_2}$.
We say in this case that $\cal C$ is a code of type $4^{k_1}2^{k_2}$.

Every quaternary code ${\cal C}$ of type $4^{k_1}2^{k_2}$
 can be defined also by a {\it check matrix}
\[A=\left[\matrix{A_1\cr 2A_2}\right]\label{pro}\]
by condition
\[Ac^T=0 \mbox{\ \ \ for all\ \ \ }c\in{\cal C},\]
where $A_1$ is a $Z_4$-matrix of size $(n-k_1-k_2)\times n$ and
$A_2$ is a $Z_2$-matrix of size $k_2\times n$. The code
${\cal C}^*$ with generator matrix $A$
is called the {\it dual} to $\cal C$.

Let two maps $\beta(c),\gamma(c):Z_4 \to Z_2$ be defined by
\[\matrix{c & \beta(c) & \gamma(c) \cr
          0&0&0\cr
          1&0&1\cr
          2&1&1\cr
          3&1&0},\]
and let they be extended coordinate-wise to maps from $Z_4^n$ to $Z_2^n$.
The {\it Gray map} $\phi: Z_4^n\to E^{2n}$ is defined by (cf.\cite{Z4})
\[\phi(c)=(\beta(c),\gamma(c)),\quad c\in Z_4^n.\]
So the $i$-th coordinate of a word $c\in Z_4^n$
corresponds to $i$-th and $(i+n)$-th
coordinates of the binary word $\phi(c)$.
In such a manner a binary code of length $2n$
corresponds to any quaternary code of length $n$.
A binary code $C$ of length $2n$ is called {\it $Z_4$-linear}
if there exist a quaternary code $\cal C$ and a permutation $\pi$ of
$2n$ coordinate such that $C=\pi(\phi({\cal C}))$.

Two binary codes $C$ and $C'$ of length $N$ are called {\it equivalent}
if there exist a word $y\in E^N$ and a permutation $\pi$
of order $N$ such that $C=\pi(C'\oplus y)$.
If quaternary codes $\cal C$ and ${\cal C}'$ are equivalent,
then related binary codes
$\phi({\cal C})$ and $\phi({\cal C}')$ are also equivalent.

The following lemma follows immediately
from definitions of distances $d(\cdot,\cdot)$,
 $d_L(\cdot,\cdot)$ and the mapping $\phi(\cdot)$

\begin{lem} {\rm \cite{Z4}} \label{thgr}
The mapping $\phi$ is an isometry from
$Z_4^n$ with Lee distance to $E^{2n}$ with Hamming distance. In other
words
\[d_L(a,b)=d(\phi(a),\phi(b)),\quad a,b\in Z_4^n.\]
\end{lem}


\section{Construction}

Let $r_1$ and $r_2$ be nonnegative integers.
Let the matrix $A^{r_1,r_2}$ consist of lexicographically ordered
columns $z^T$,
$z\in \{1\}\times\{0,1,2,3\}^{r_1}\times\{0,2\}^{r_2}$.
For example
\[A^{0,0}=\left[\matrix{1}\right],\quad
    A^{0,1}=\left[\matrix{11\cr02}\right],\]
\[A^{1,0}=\left[\matrix{1111\cr0123}\right], \quad
    A^{0,2}=\left[\matrix{1111\cr0022\cr0202}\right],\]
\[A^{1,1}=\left[\matrix{11\,11\,11\,11\cr00\,11\,22\,33\cr02\,02\,02\,02}\right], \quad
    A^{0,3}=\left[\matrix{11\,11\,11\,11\cr
                          00\,00\,22\,22\cr
                          00\,22\,00\,22\cr
                          02\,02\,02\,02}\right],\]
\[A^{2,0}=\left[\matrix{1111\,1111\,1111\,1111\cr
                           0000\,1111\,2222\,3333\cr
                           0123\,0123\,0123\,0123}\right].\]

For all integers $r_1,r_2\geq 0$ define the dual
quaternary codes ${\cal H}^{r_1,r_2}$ and ${\cal C}^{r_1,r_2}$:
\[{\cal H}^{r_1,r_2}\stackrel{\rm\scriptscriptstyle def}=\{(v_1,v_2)A^{r_1,r_2}:v_1\in Z_4^{r_1+1}, v_2\in Z_2^{r_2} \},\]
\[{\cal C}^{r_1,r_2}\stackrel{\rm\scriptscriptstyle def}=\{c\in Z_4^{2^{2r_1+r_2}}:A^{r_1,r_2}c^T=0\}.\]
The matrix $A^{r_1,r_2}$ is
a generator matrix for ${\cal H}^{r_1,r_2}$
and a check matrix for ${\cal C}^{r_1,r_2}$.

Let $n=2^{2r_1+r_2}$.

\begin{thm}\label{th4_1}
{\rm a)} The set ${\cal H}^{r_1,r_2}$ is a quaternary $(n,    4n,n)_4$-code$;$
\\[0.15ex]
{\rm b)} the set ${\cal C}^{r_1,r_2}$ is a quaternary $(n,4^n/4n,4)_4$-code.
\end{thm}

Let $H^{r_1,r_2}\stackrel{\rm\scriptscriptstyle def}=\phi({\cal H}^{r_1,r_2})$,
$C^{r_1,r_2}\stackrel{\rm\scriptscriptstyle def}=\phi({\cal C}^{r_1,r_2})$ and
let $N\stackrel{\rm\scriptscriptstyle def}=2n=2^{2r_1+r_2+1}$. By Lemma~\ref{thgr}, Theorem~\ref{th4_1}
means that

\begin{cor}\label{crll4_1}
{\rm a)} The set $H^{r_1,r_2}$ is a binary $(N,2N,N/2)$-code$;$
\\[0.15ex]
{\rm b)} the set $C^{r_1,r_2}$ is a binary $(N,2^N/2N,4)$-code.
\end{cor}


\section{The nonexistence of $(n,4n,n)_4$- and $(n,4^n/4n,4)_4$-codes that are nonequivalent to the constructed codes}

\begin{thm}\label{th4_3}
{\rm a)} Let the set ${\cal H}\subset Z_4^n$ be a $(n,4n,n)_4$-code
of type $4^{r_0}2^{r_2}$. Then
$n=2^{2(r_0-1)+r_2}$,
$r_0>0$,
and  $\cal H$ is equivalent to ${\cal H}^{r_0-1,r_2}$.
\\[0.15ex]
{\rm b)} Let the set ${\cal C}\subset Z_4^n$ be a $(n,4^n/4n,4)_4$-code
of type $4^{n-r_0-r_2}2^{r_2}$. Then
$n=2^{2(r_0-1)+r_2}$,
$r_0>0$,
and $\cal C$ is equivalent to ${\cal C}^{r_0-1,r_2}$.
\end{thm}

\begin{cor}\label{col}
{\rm a)} Each $Z_4$-linear $(N,2N,N/2)$-code is equivalent to some
code $H^{r_1,r_2}$, $2^{2r_1+r_2+1}=N$.
\\[0.15ex]
{\rm b)} Each $Z_4$-linear $(N,2^N/2N,4)$-code is equivalent to
some $C^{r_1,r_2}$, $2^{2r_1+r_2+1}=N$.
\end{cor}


\section{The nonequivalence of $H^{r_1,r_2}$}

If $H$ is a binary code of length $N$ then
\[{\rm kernel}(H)\stackrel{\rm\scriptscriptstyle def}=\{x\in E^N : x\oplus H=H\}.\]

The proof of pairwise nonequivalency of the codes $H^{r_1,r_2}$
is based on the following fact.

\begin{prop}
\label{p1} If binary codes $H_1$ and $H_2$ are equivalent then
 $|{\rm kernel}(H_1)|=|{\rm kernel}(H_2)|$.
\end{prop}

The following two propositions establish the cardinalities of kernels
of the codes $H^{r_1,r_2}$.

\begin{prop} The codes $H^{0,r_2}$ and $H^{1,r_2}$ are linear.
Hence ${\rm kernel}(H^{0,r_2})=H^{0,r_2}$ and ${\rm kernel}(H^{1,r_2})=H^{1,r_2}$.
\end{prop}

\begin{prop} \label{p3} Let $r_1>1$. Then $|{\rm kernel}(H^{r_1,r_2})|=2^{r_1+r_2+2}$ and
the code $H^{r_1,r_2}$ is nonlinear.
\end{prop}

The following theorem stems from Propositions \ref{p1}-\ref{p3}.

\begin{thm}\label{ththth} Let $2r_1+r_2=2r'_1+r'_2$ and $r_1\geq 2$.
Then the codes $H^{r_1,r_2}$ and
$H^{r'_1,r'_2}$ are equivalent if and only if $r_1=r'_1$.
\end{thm}

From Theorem \ref{ththth},
using Corollary \ref{col}, we have

\begin{thm}\label{z4main_ad}
If $N=2^k\geq 8$ then there exist exactly $\lfloor(k-1)/2\rfloor$
pairwise nonequivalent $Z_4$-linear Hadamard codes of length $N$.
\end{thm}


\section{The nonequivalence of $C^{r_1,r_2}$}

If $N$ is even and $C\subset E^N$ then
\[{\rm even}(C)\stackrel{\rm\scriptscriptstyle def}=\{(c_0,c_2,\ldots,c_{N-2})\in E^{N/2}\,|\,
(c_0,0,c_2,0,\ldots,c_{N-2},0)\in C\},\]
\[{\rm odd}(C)\stackrel{\rm\scriptscriptstyle def}=
\{(c_1,c_3,\ldots,c_{N-1})\in E^{N/2}\,|\,
(0,c_1,0,c_3,\ldots,0,c_{N-1})\in C\}.\] We use these definitions
and the following proposition for the induction step.

\begin{prop}\label{CCC} It is true that\\[0.15ex]
{\rm a)}
${\rm even}(C^{r_1,r_2})=
{\rm odd}(C^{r_1,r_2})=C^{r_1,r_2-1}$ for every $r_1\geq 0$ and $r_2>0$;
\\[0.15ex]
{\rm b)}
${\rm even}(C^{r_1,0})={\rm odd}(C^{r_1,0})=C^{r_1-1,1}$ for every $r_1>0$.
\end{prop}

Let the maximal
number of linearly independent vectors from a binary code $C$
be noted ${\rm rank}(C)$.

The proof of pairwise nonequivalence of $C^{r_1,r_2}$
is based on the following fact.

\begin{prop} If binary codes $C_1$ and $C_2$ are equivalent
then ${\rm rank}(C_1)={\rm rank}(C_2)$.
\end{prop}

\begin{prop}\label{r1r2}
For all integers $r_1\geq 0$, $r_2\geq 0$
\begin{equation}{{\rm rank}(C^{r_1,r_2})\leq N-r_1-r_2-1,\label{nerav}}
\end{equation}
where $N=2^{2r_1+r_2+1}$ is the length of code $C^{r_1,r_2}$.
\end{prop}

It is straightforward that (\ref{nerav})
is tight for $r_1=r_2=1$ and $r_1=0$, $r_2=4$:

\begin{prop}\label{C16} It is true that ${\rm rank}(C^{1,1})=13$ and
${\rm rank}(C^{0,4})={27}$.
\end{prop}

Using Proposition~\ref{CCC} it can be established by induction
that (\ref{nerav}) is tight for every $r_1,r_2\geq 1$ or
$r_1\geq 0,\,r_2\geq 4$:

\begin{lem}\label{rank} Let $r_1\geq 1$, $r_2\geq 0$ be integers such that
$2r_1+r_2\geq 3$ and $(r_1,r_2)\neq (0,3)$.
Then \[{\rm rank}(C^{r_1,r_2})=2^{2r_1+r_2+1}-r_1-r_2-1.\]
\end{lem}

\begin{rem} The set $C^{0,3}$ is a linear code and ${\rm rank}(C^{0,3})=11$.
\end{rem}

\begin{thm}\label{th4_2} Let $2r_1+r_2=2r'_1+r'_2\geq 3$. Then
the codes $C^{r_1,r_2}$ and $C^{r'_1,r'_2}$ are equivalent if and only if
 $r_1=r'_1$ $($equivalently, $r_2=r'_2)$.
\end{thm}

By Corollary~\ref{crll4_1} and Corollary~\ref{col} we have

\begin{thm}\label{z4main_hm}
If $N=2^k\geq 16$ then there exist exactly $\lfloor(k+1)/2\rfloor$
pairwise nonequivalent $Z_4$-linear
extended perfect distance $4$ codes of length $N$.
\end{thm}


\section{Recurrent construction of codes ${\cal H}^{r_1,r_2}$}
Let $\cal H$ be a quaternary $(n,4n,n)_4$-code, ${\cal R}'=\{00...0,22...2\}$ be
the quaternary $(n,2,2n)_4$-code, and ${\cal R}''=\{00...0,11...1,22...2,33...3\}$ be
the quaternary repetition $(n,4,n)_4$-code. Let
\begin{equation} H'\stackrel{\rm\scriptscriptstyle def}=\{(a,a+b):a\in {\cal H}, b\in {\cal R}'\}, \label{Plotk}
\end{equation}
\[ H''\stackrel{\rm\scriptscriptstyle def}=\{(a,a+b,a+2b,a+3b):a\in {\cal H}, b\in {\cal R}''\}.\]

\begin{rem} The construction (\ref{Plotk}) is a particular case of
well known Plotkin $(u,u+v)$-construction.
\end{rem}

\begin{prop}\label{pr8}
The set ${\cal H}'$ is a quaternary $(2n,4(2n),2n)_4$-code.
If ${\cal H}$ is equivalent to ${\cal H}^{r_1,r_2}$ then ${\cal H}'$
is equivalent to ${\cal H}^{r_1,r_2+1}$.
If ${\cal H}={\cal H}^{0,r_2}$ then ${\cal H}'={\cal H}^{0,r_2+1}$.
\end{prop}

\begin{prop}\label{pr9}
The set ${\cal H}''$ is a quaternary $(4n,4(4n),4n)_4$-code.
If ${\cal H}={\cal H}^{r_1,r_2}$ then ${\cal H}'={\cal H}^{r_1+1,r_2}$.
\end{prop}

Using Propositions~\ref{pr8} and~\ref{pr9}
one can construct every code ${\cal H}^{r_1,r_2}$
starting with the trivial code ${\cal H}^{0,0}=\{0,1,2,3\}$.

A recurrent construction of the class of codes ${\cal C}^{r_1,r_2}$ can be found in \cite{Kr}.

\end{document}